\documentclass[twocolumn]{aastex62}

\usepackage{float}
\usepackage{hyperref}
\usepackage{amsmath}
\hypersetup{colorlinks=true,linkcolor=cyan,citecolor=cyan,filecolor=cyan,urlcolor=cyan,}

\newcommand{\mstel}{\mathrm{M_*}}
\newcommand{\msun}{\mathrm{M_\odot}}

\newcommand{\ha}{H$\mathrm{\alpha}$}

\newcommand{\dnfour}{$\mathrm{D_n4000}$}
\newcommand{\dhost}{$\mathrm{d}_\mathrm{{host}}$}
\newcommand{\angstrom}{\mbox{\normalfont\AA}}

\graphicspath{{./}{figures/}}

\shorttitle{IQ II: Low Mass Quiescent Fractions}
\shortauthors{Dickey et al.}

\begin{document}

\title{IQ Collaboratory II: The Quiescent Fraction of Isolated, Low Mass Galaxies Across Simulations and Observations}

\correspondingauthor{Claire Dickey}
\email{claire.dickey@yale.edu}

\author[0000-0002-1081-3991]{Claire M. Dickey}
\affiliation{Department of Astronomy, Yale University, 52 Hillhouse Ave, New Haven, CT 06520, USA}

\author[0000-0003-2539-8206]{Tjitske K. Starkenburg}
\affiliation{Center for Computational Astrophysics, Flatiron Institute, 162 Fifth Avenue, New York, NY 10010, USA}
\affiliation{Center for Interdisciplinary Exploration and Research in Astrophysics (CIERA) and\\ Department of Physics and Astronomy, Northwestern University, 1800 Sherman Ave, Evanston IL 60201, USA}

\author[0000-0002-7007-9725]{Marla Geha}
\affiliation{Department of Astronomy, Yale University, 52 Hillhouse Ave, New Haven, CT 06520, USA}

\author{ChangHoon Hahn}
\affil{Lawrence Berkeley National Laboratory, 1 Cyclotron Rd, Berkeley CA 94720, USA}
\affil{Berkeley Center for Cosmological Physics, University of California, Berkeley, CA 94720, USA}
\affil{Department of Astrophysical Sciences, Princeton University, Peyton Hall, Princeton NJ 08544, USA}

\author{Daniel Angl\'{e}s-Alc\'{a}zar}
\affiliation{Department of Physics, University of Connecticut, 196 Auditorium Road, U-3046, Storrs, CT 06269-3046, USA}
\affiliation{Center for Computational Astrophysics, Flatiron Institute, 162 Fifth Avenue, New York, NY 10010, USA}

\author[0000-0002-8131-6378]{Ena Choi}
\affiliation{Quantum Universe Center,
        Korea Institute for Advanced Study,
        Hoegiro 85, Seoul 02455, Korea}

\author{Romeel Dav\'{e}}
\affiliation{Institute for Astronomy, Royal Observatory, Univ. of Edinburgh, Edinburgh EH9 3HJ, UK}
\affiliation{University of the Western Cape, Bellville, Cape Town 7535, South Africa}
\affiliation{South African Astronomical Observatories, Observatory, Cape Town 7925, South Africa}

\author{Shy Genel}
\affiliation{Center for Computational Astrophysics, Flatiron Institute, 162 Fifth Avenue, New York, NY 10010, USA}
\affiliation{Columbia Astrophysics Laboratory, Columbia University, 550 West 120th Street, New York, NY 10027, USA}

\author[0000-0001-9298-3523]{Kartheik G. Iyer}
\affiliation{Dunlap Institute for Astronomy and Astrophysics, University of Toronto, 50 St George St, Toronto, ON M5S 3H4, Canada}

\author[0000-0003-2060-8331]{Ariyeh H. Maller}
\affiliation{Department of Physics, New York City College of Technology,
City University of New York, 300 Jay St., Brooklyn, NY 11201, USA}

\author{Nir Mandelker}
\affiliation{Department of Astronomy, Yale University, 52 Hillhouse Ave, New Haven, CT 06520, USA}
\affiliation{Heidelberger Institut f\"{u}r Theoretische Studien, Schloss-Wolfsbrunnenweg 35, 69118 Heidelberg, Germany}
\affiliation{Kavli Institute for Theoretical Physics, University of California, Santa Barbara, CA 93106, USA}
\affiliation{Racah Institute of Physics, The Hebrew University of Jerusalem, Jerusalem 91904, Israel}

\author{Rachel S. Somerville}
\affiliation{Center for Computational Astrophysics, Flatiron Institute, 162 Fifth Avenue, New York, NY 10010, USA}
\affiliation{Department of Physics and Astronomy, Rutgers University, 136 Frelinghuysen Road, Piscataway, NJ 08854}

\author{L. Y. Aaron Yung}
\affiliation{Department of Physics and Astronomy, Rutgers University, 136 Frelinghuysen Road, Piscataway, NJ 08854}
\affiliation{Center for Computational Astrophysics, Flatiron Institute, 162 Fifth Avenue, New York, NY 10010, USA}
\affiliation{ Astrophysics Science Division, NASA Goddard Space Flight Center, 8800 Greenbelt Rd, Greenbelt, MD 20771, USA}


\begin{abstract}

    We compare three major large-scale hydrodynamical galaxy simulations (EAGLE, Illustris-TNG, and SIMBA) by forward modeling simulated galaxies into observational space and computing the fraction of isolated and quiescent low mass galaxies as a function of stellar mass. Using SDSS as our observational template, we create mock surveys and synthetic spectroscopic and photometric observations of each simulation, adding realistic noise and observational limits. All three simulations show a decrease in the number of quiescent, isolated galaxies in the mass range $\mstel = 10^{9-10} \ \msun$, in broad agreement with observations. However, even after accounting for observational and selection biases, none of the simulations reproduce the observed absence of quiescent field galaxies below $\mstel=10^{9} \ \msun$. We find that the low mass quiescent populations selected via synthetic observations have consistent quenching timescales, despite apparent variation in the late time star formation histories. The effect of increased numerical resolution is not uniform across simulations and cannot fully mitigate the differences between the simulations and the observations. The framework presented here demonstrates a path towards more robust and accurate comparisons between theoretical simulations and galaxy survey observations, while the quenching threshold serves as a sensitive probe of feedback implementations.

\end{abstract}



\section{Introduction} \label{sec:intro}

    The transition of galaxies from star-forming into quiescence provides key insights into the physical processes that drive galaxy evolution. Major extragalactic surveys like the Sloan Digital Sky Survey \citep[SDSS, ][]{york2000} have been instrumental in quantifying the differing properties of galaxies in these two categories, including morphology, environment, and color \citep[e.g.,][]{blanton2003,kauffmann2003,brinchmann2004,blanton2009,moustakas2013}. 
    The processes that drive a galaxy's transformation can be broadly divided into two categories: external processes which occur in high-density environments and processes internal to the galaxy which are often thought to be correlated to galaxy mass \citep[e.g.,][]{peng2010}.
    
    Understanding how and when feedback mechanisms operate within galaxies and in which regimes they are most effective remains a major focus of both observational and theoretical studies of galaxy evolution. Observational galaxy surveys have been used to characterize the populations of quiescent and star forming galaxies as a function of stellar mass \citep{kauffmann2004,geha2012}, environment \citep{wetzel2012}, and redshift \citep{brammer2009}; as well as key correlations between galaxy properties \citep[e.g., the star forming sequence; ][]{daddi2007,noeske2007,salim2007,elbaz2007} which inform our understanding of the underlying forces that drive galaxy evolution.
    
    Large-scale cosmological hydrodynamic galaxy formation simulations reproduce the observed universe in a qualitative manner \citep{genel2014,vogelsberger2014,schaye2015,dave2017,pillepich2018b, dave2019}. Each simulation uses a distinct set of approximations for the complex physics of underlying galaxy evolution, including star formation, heating and cooling of gas, black hole formation and growth, feedback from active galactic nuclei (AGN), and stellar feedback \citep[for an overview see][]{somerville2015,vogelsberger2020}. While it has been shown that many different subgrid models can produce relatively consistent pictures of galaxy evolution \citep{naab2017}, the variations between simulations present a unique opportunity to explore how feedback mechanisms shape galaxy evolution.
    
    Many studies have focused on comparing the simulated distributions of quantities such as galaxy masses, colors, or star-formation rates to observations \citep[e.g.,][]{genel2014, torrey2014, vogelsberger2014, schaye2015, somerville2015, sparre2015, dave2017, pillepich2018b, nelson2018, trayford2017, donnari2019,donnari2020_obs,donnari2020}. These works primarily focus on comparing a singular simulation to a particular set of observations. Studies using multiple simulations require the careful construction of a consistent framework for inter-simulation comparison.
    
    In the observed universe, the quiescent fraction for isolated galaxies is zero at $\mstel < 10^9 \ \msun$ \citep{geha2012} in the SDSS volume, suggesting that feedback in this mass regime is highly sensitive to stellar mass. Most studies of feedback in low mass galaxies have focused on environmental quenching, either in the context of high density environments such as clusters or interactions with a Milky Way-like host galaxy \citep[e.g.,][]{fillingham2018, sales2015, wetzel2013, wetzel2015}.
    
    Internal feedback may influence galaxy evolution down to $\mstel \sim 10^9 \ \msun$, via either AGN \citep{dickey2019,koudmani2019,penny2018} or stellar feedback \citep{elbadry2016}. However, the efficiency and interplay of different feedback mechanisms below this mass limit remains uncertain. Large-scale hydrodynamic simulations present a unique opportunity to create an ``observed'' magnitude limited galaxy survey and explore how different subgrid feedback implementations shape the distribution.
    
    The IQ (Isolated \& Quiescent) Collaboratory\footnote{https://iqcollaboratory.github.io/} aims to bridge the gap between simulations and observations of star-forming and quiescent galaxies to better characterize internal quenching processes. In \citet[][Paper I]{IQ_paper1}, we began the process of comparing the star forming sequence across a set of simulations. 
        
    Our initial analysis in Paper I highlighted the large variation in the apparent quiescent populations of each simulation, but we did not make a direct comparison between simulations and observations. In that work, the star formation rates from the simulations are obtained ``directly'' from the simulation output, as opposed to being derived from ``observables'' like \ha- or UV+IR-derived star formation rates, or indices such as the \ha\ equivalent width (EW) and \dnfour, an index that measures the strength of the $4000 \angstrom$ break and loosely traces stellar age. Moreover, even within observational results, selecting different tracers for e.g., SFRs can give rise to significant variation \citep[e.g.,][]{salim2007,kennicutt2012, speagle2014, flores2020, katsianis2020}
    
    In addition to deriving galaxy observables from a simulation, fully transforming a simulated volume into a true  observational analogue requires the careful creation of complete, volume-corrected samples. This is particularly challenging when we are trying to reproduce observations comparing star forming and quiescent populations, which often suffer from differing levels of incompleteness within the same survey.
    
    Our goal in this work is to create an ``apples-to-apples'' comparison of the quiescent  fraction of isolated low mass ($\mstel < 10^{10} \ \msun$) galaxies as a function of stellar mass, using observations of the local universe and a set of simulations, all of which produce realistic cosmological volumes using distinct implementations of subgrid physics. We focus our study on a selection of three major cosmological hydrodynamic simulations (EAGLE, Illustris-TNG, and SIMBA) and compare them to the observed population of low mass quiescent galaxies in the Sloan Digital Sky Survey (SDSS). 
    
    In \autoref{sec:sims}, we present the simulations used in this work and in \autoref{sec:obs} we discuss the observations which serve as our point of comparison. In \autoref{sec:mock_obs}, we create mock observations from each simulation, including synthetic spectra, realistic noise, and SDSS-like incompleteness. \autoref{sec:obs_v_sims} compares the ``observed'' simulations to the distribution of low mass quiescent galaxies in SDSS. In \autoref{sec:SFH} we explore the star formation histories of quiescent galaxies in the simulations and we review our findings in \autoref{sec:conc}.


\section{Galaxy Formation Simulations} \label{sec:sims}

    In this work, we focus on three large-scale hydrodynamic cosmological simulations (EAGLE, Illustris-TNG, and SIMBA), which we will compare to observations. We provide brief descriptions of each simulation below.
    
    \subsection{EAGLE}
    
        The Virgo Consortium's Evolution and Assembly of GaLaxies and their Environment (EAGLE) project\footnote{https://icc.dur.ac.uk/Eagle/} \citep{schaye2015,crain2015,mcalpine2016} has a volume of (100 Mpc)$^3$ (co-moving), with dark matter and baryonic particle masses of $9.6 \times 10^6 \ \msun$ and $1.8 \times 10^6 \ \msun$. The simulation uses ANARCHY (Dalla Vecchia, in prep.), a modified version of the Gadget3 N-body SPH code \citep{springel2001b,springel2005, schaller2015}. The subgrid model for feedback from massive stars and AGN is based on thermal energy injection in the interstellar medium \citep{dalla2012}. The subgrid parameters were calibrated to reproduce the $z = 0$ stellar mass function and galaxy sizes.  
        
        Previous works that have reproduced observables and examined the quenched population of the EAGLE simulation include \citet{schaye2015}, \citet{furlong2015}, \citet{trayford2015} and \citet{trayford2017}, and the galaxy--black hole relations are discussed in \citet{mcalpine2017, mcalpine2018, bower2017, habouzit2020}. \citet{trcka2020} show that mock spectral energy distributions (see \citealp{camps2018} for public release of the SEDs) based on EAGLE galaxies and radiative transfer calculations using \texttt{skirt} \citep{baes2011} are in overall agreement with observed galaxy spectra, and highlight the need for consistent comparisons between simulations and observations. \citet{trayford2017} used the same results of the \texttt{skirt} radiative transfer code to determine quenched fractions based on mock UVJ colors, along with the distribution of \ha\ flux and \dnfour. They found that the passive fraction varies significantly depending on the definition of quenched. We will compare our results to previous conclusions in \autoref{subsec:EAG_disc}.
    
    \subsection{Illustris-TNG}
    
        \textit{The Next Generation} Illustris project (IllustrisTNG or TNG)\footnote{https://www.tng-project.org/} \citep{marinacci2018, naiman2018, nelson2018, pillepich2018b, springel2018} is the successor to the original Illustris project \citep{genel2014, vogelsberger2014}, with significant updates in the subgrid models and physics included in the simulation. We use the TNG100 simulation, which has a volume of (110.7 Mpc)$^3$, and dark matter and baryonic mass resolutions of $7.6 \times 10^6 \ \msun$ and $1.4 \times 10^6 \ \msun$. TNG is run using the AREPO moving-mesh code \citep{springel2010}, which is based on the Gadget code \citep{springel2001b, springel2005}. Adjustments in the TNG model include the addition of magneto-hydrodynamics, updated stellar feedback prescriptions, and the transition from thermal ``bubbles'' in the IGM to a BH-driven kinetic wind for the low-accretion-rate black hole feedback mode \citep{pillepich2018a, weinberger2017}. 
        
        The color bimodality of low-redshift galaxies is shown to compare well to SDSS \citep{nelson2018}. Other papers describe the size evolution of quiescent galaxies \citep{genel2018}, and correlations between galaxy properties and super massive black holes and AGN feedback \citep{weinberger2018, habouzit2019, terrazas2020, davies2020, li2019, habouzit2020}. \citet{donnari2019} show quenched fractions based on UVJ selection and the star forming sequence, and compare these to UVJ-selected observed quenched fractions from COSMOS/UltraVISTA \citep{muzzin2013}. \citet{donnari2020_obs} and \citet{donnari2020} explore the quenched fraction derived using star formation rates and distances from the galaxy star-forming sequence.  \citet{donnari2020_obs} in particular explores how systematic uncertainties effect a single set of observations, which we will further discuss in \autoref{subsec:TNG_disc}.

    \subsection{SIMBA}
    
        SIMBA \citep{dave2019} is a suite of cosmological simulations built on the \texttt{GIZMO} meshless finite mass hydrodynamics code \citep{hopkins2015, hopkins2017}, also based on the Gadget code \citep{springel2001b, springel2005}, and forms the next generation of the MUFASA \citep{dave2016} simulations with novel black hole growth and feedback sub-grid models. The fiducial run has a volume of $(143 \ \mathrm{Mpc})^3$ and dark matter and baryonic mass resolutions of $9.6 \times 10^7 \ \msun$ and $1.8 \times 10^7 \ \msun$, respectively. 
        
        SIMBA includes a model for on-the-fly dust production and destruction \citep[broadly following][]{mckinnon2017}, and star formation is regulated with two-phase kinetic outflows, which were tuned to predictions from the Feedback in Realistic Environments (FIRE) simulations \citep{hopkins2014, muratov2015, angles-alcazar2017b, hopkins2018}. Black hole growth in SIMBA is based on the torque-limited accretion model \citep{hopkins2011, angles-alcazar2013, angles-alcazar2015, angles-alcazar2017} linking the black hole accretion to properties of the galaxy's inner gas disk. The AGN feedback consists of kinetic bipolar outflows, modeled after observed outflows of AGN, and X-ray feedback input in the surrounding gas similar to \citet{choi2012}. 
        
        Previous work using SIMBA has already discussed the radial density profiles of quenched galaxies \citep{appleby2019}, the weak correlation between galaxy mergers and quenching \citep{rodriguez-montero2019}, and the connection between quenching and black hole growth \citep{dave2019, thomas2019, habouzit2020}. We compare our results to relevant studies in \autoref{subsec:SIMBA_disc}.

\section{Observations}\label{sec:obs}
    
    Following \citet[][]{geha2012}, we build our sample of observed galaxies from SDSS, using the NASA/Sloan Atlas \citep[NSA;][]{blanton2011}\footnote{https://www.nsatlas.org}, which includes all galaxies within $z < 0.055$ in the SDSS footprint. The NASA/Sloan Atlas is a re-reduction of SDSS DR8 \citep{aihara2011} optimized for nearby low-luminosity objects \citep{yan2011,yan2012}, including an improved background subtraction technique \citep{blanton2011} and the addition of near and far UV photometry from \textit{GALEX}. The catalog includes emission line fluxes and equivalent widths for all galaxies. We calculate stellar masses as in \citet{mao2020}, using the \citet{bell2003} relation to determine mass-to-light ratio from the $g-r$ color. See \autoref{subsec:mass} for a more in-depth discussion on the effects of uncertainty in the stellar mass measurements on the resultant quiescent fraction.
    
    As in \citet{geha2012}, we consider a low mass galaxy to be quiescent if it has little-to-no star formation and is dominated by older stellar populations. For the first condition, we use the \ha\ equivalent width (EW), which traces recent ($< 10$ Myr) star formation, and require \ha\ EW $< 2$ \AA. To probe galaxy age, we rely on \dnfour\ \citep{balogh1999}, an index which quantifies the strength of the $4000\,$\AA\ break in the spectrum and traces the light-weighted age of the stellar population. The \dnfour\ index is an indirect measure of intermediate (${\sim}1$ Gyr) star formation \citep{hamilton1985, moustakas2006, brinchmann2004}. We use the empirical relation of \citet{geha2012}: \dnfour\ $> 0.6+0.1 \log_{10}(\mstel / \msun)$, to select galaxies with older stellar populations.

    To quantify the environment of each low mass galaxy, we use \dhost, which is defined as the projected distance to the nearest massive neighbor ($\mstel > 2.5 \times 10^{10} \ \msun$; $\mathrm{M_{K_s}} < -23$; hereafter the host galaxy). We search for potential host galaxies within 7 Mpc and 1000 km/s of each low mass galaxy, using the Two Micron All Sky Survey (2MASS) Extended Source Catalog to ensure we do not miss any potential host galaxies that lie outside the SDSS imaging footprint. In the few cases where we do not identify a host within 7 Mpc, we use \dhost = 7 Mpc. 
    
    We define galaxies as isolated when \dhost\ $> 1.5$ Mpc. This is an empirical choice based on the behavior of the quiescent fraction as a function of \dhost\ (see Fig. 4 of \citet{geha2012} or \autoref{fig:dhost} in this work). The quiescent fraction only shows a dependence on environment for \dhost\ $< 1.5$ Mpc, and so we consider low mass galaxies beyond this threshold to be isolated. Using a less strict definition (e.g., galaxies are isolated beyond \dhost \ $ = 1$ Mpc) shifts the quenching threshold to a slightly lower mass, while a more conservative definition does not change the observed threshold. For galaxies with stellar masses above $\mstel=10^{10} \ \msun$, we use the central -- satellite designation from the group catalog of \citet{tinker2011}. 
    
\begin{figure*}
        \epsscale{1.2}
            \plotone{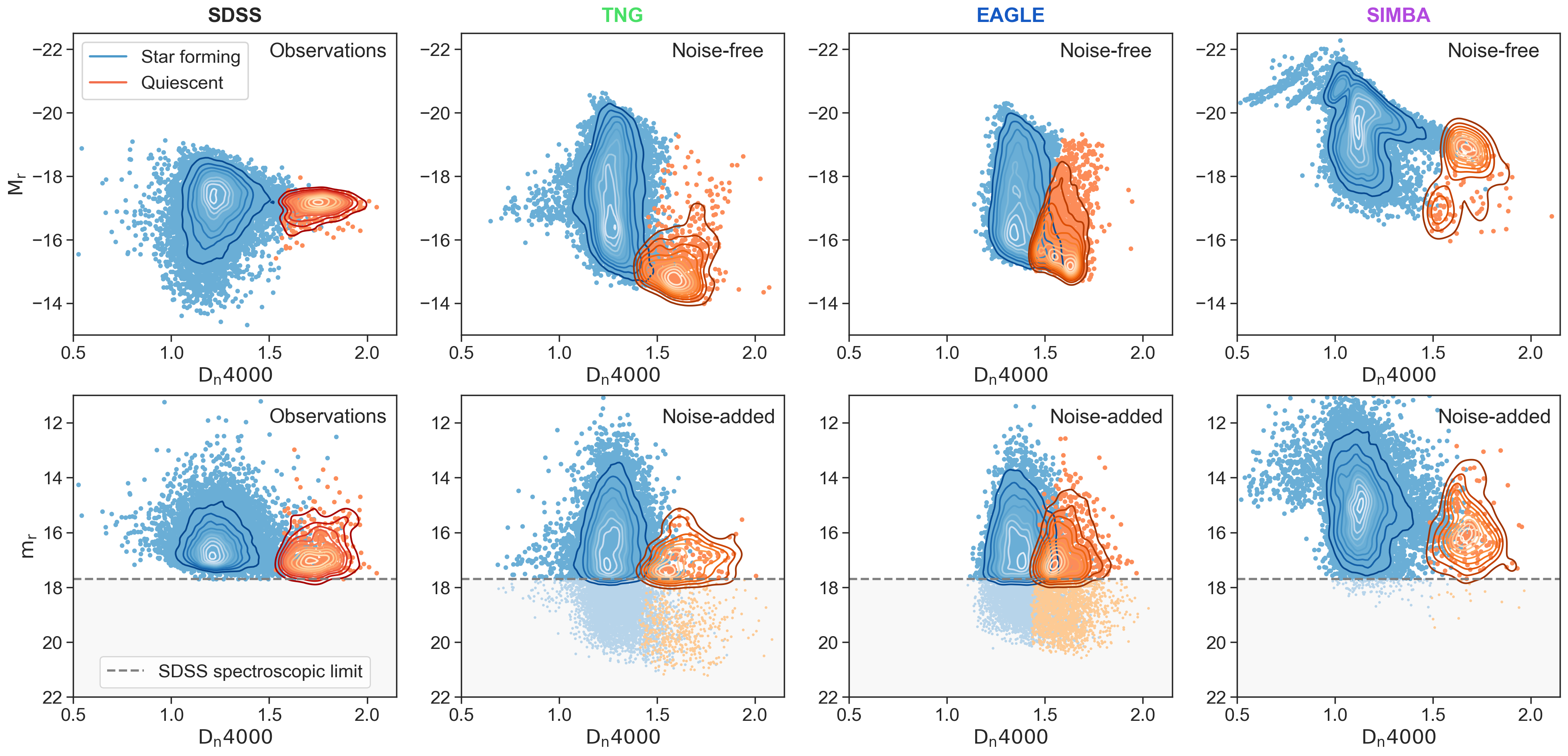}
            \caption{\textbf{\textit{Top row:}} The distribution of absolute $r$ magnitude ($\mathrm{M_r}$) as a function of \dnfour, from SDSS for observations and derived from the noise-less synthetic spectra from each simulation (from left to right: TNG, EAGLE, and SIMBA), for isolated galaxies (based on \dhost) in the stellar mass range $\mstel = 10^{8-10} \ \msun$. Galaxies are color-coded as star forming (blue) or quiescent (orange) based on the noise-free \ha\ EW and \dnfour.
            \textbf{\textit{Bottom row:}} The distribution of apparent $r$ magnitude ($\mathrm{m_r}$) as a function of \dnfour, derived from the noise-added synthetic spectra along a single random sightline through each simulation box. Galaxies are color-coded as star forming (blue) or quiescent (orange) based on the noise-added \ha\ EW and \dnfour. Galaxies below the gray dashed line fall below the SDSS spectroscopic limit and would not be selected for spectroscopic follow-up in SDSS. As such, they are not included in the calculation of $f_q$ along a given sightline.}
            \label{fig:mag_d400}
    \end{figure*}

\begin{figure*}
        \epsscale{1.2}
            \plotone{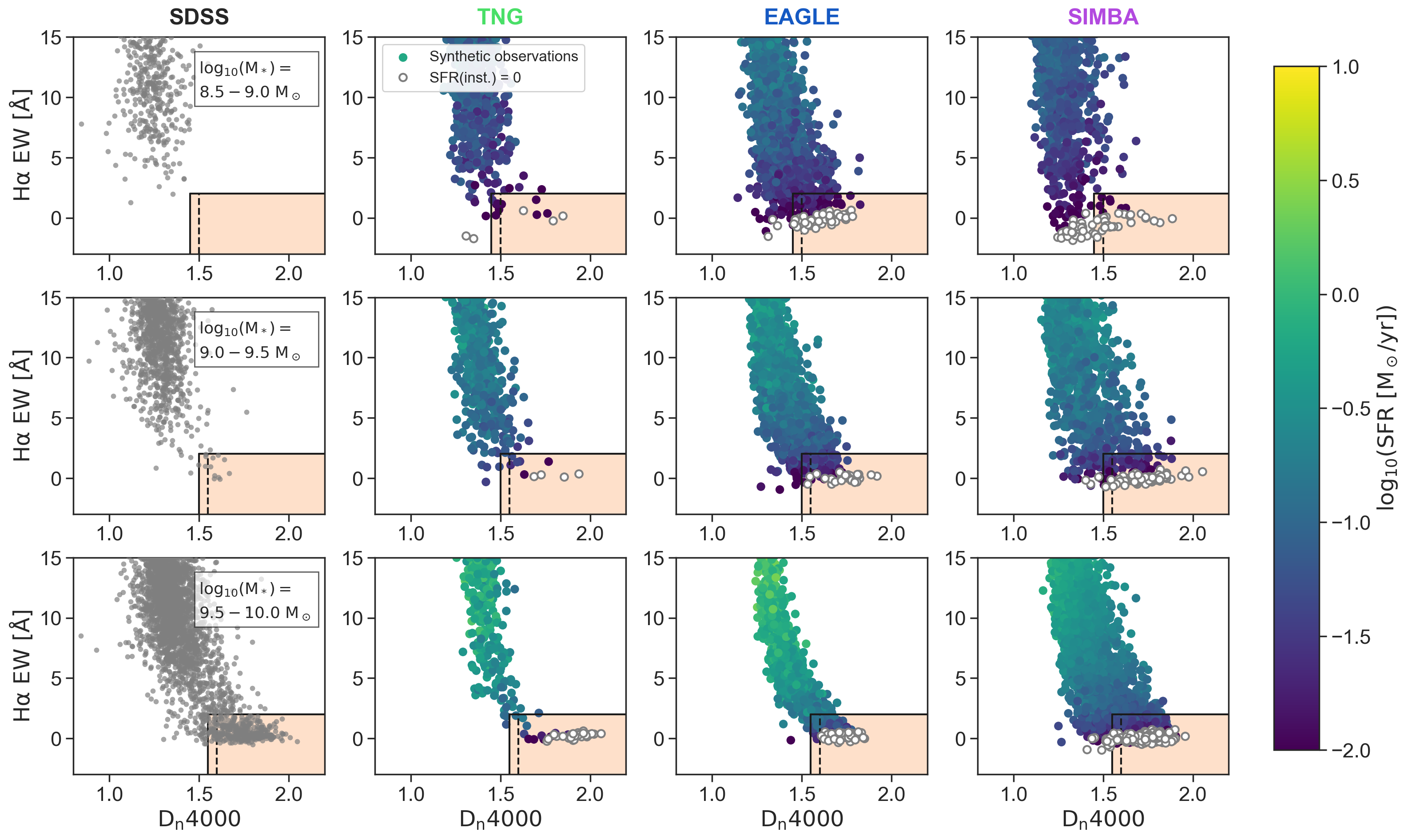}
            \caption{\dnfour\ vs.\ \ha\ EW, from SDSS for observations and derived from the noise-added synthetic spectra from each simulation (from left to right: TNG, EAGLE, and SIMBA), for isolated galaxies (based on \dhost) in three stellar mass bins (from top to bottom: $\mstel = 10^{8.5-9.0} \ \msun$, $10^{9.0-9.5} \ \msun$, and $10^{9.5-10.0} \ \msun$). Galaxies are observed as quiescent if they fall into the orange region of \dnfour-\ha\ EW parameter space. The \dnfour\ limit depends on stellar mass and is equal to the solid (dashed) vertical line at the low mass (high-mass) end of the bin for each row. For the simulations, galaxies are color-coded by their instantaneous star formation rates, and galaxies with SFR $= 0 \ \mathrm{M_\odot / yr}$ are shown in white.}
            \label{fig:d4ha_sfr}
    \end{figure*}

    In the left column of \autoref{fig:mag_d400}, we show the distribution of isolated galaxies observed in the local universe as a function of \dnfour\ and SDSS $r$ magnitude in the mass range $\mstel = 10^{8-10} \ \msun$. Galaxies in isolation are only observed to be quiescent above $\mstel = 10^9 \ \msun$, (\autoref{fig:d4ha_sfr}). 
    
\section{Mock Observations} \label{sec:mock_obs}

    To create a true apples-to-apples comparison between the observations and simulations, we must account for the effects of observational incompleteness, finite signal to noise, and a lack of precise distance information for individual galaxies in a wide-field survey. 
    
    To that end, we create mock surveys and synthetic observations of each simulation, adding realistic noise and observational limits. We use SDSS as our observational template. We select the mock observational limits and injected noise to match SDSS. We apply the same methodology to all the simulations, as described below for a single simulation box. However, this method can also be generalized and applied to semi-analytic models and zoom-in simulations, as well as adapted to match other surveys and observations.
    
    \subsection{Synthetic spectra}\label{subsec:mock_spectra}
    
        We begin by generating synthetic spectra for all galaxies in a given simulation using \texttt{FSPS} (Flexible Stellar Population Synthesis; \citealp{conroy2009,conroy2010}, and the Python interface \texttt{python-fsps} \citealp{foremanmackey}). Our method is described in full in Starkenburg et.\ al (in prep.), but in brief it is as follows.
        
        For each galaxy, we bin the total stellar mass formed by formation age ($t$) and stellar metallicity ($Z$). We use a uniform $t,Z$ grid across all the simulations to standardize the effects of otherwise variable time resolution and particle size. Age bin size increases linearly as a function of lookback time based on the minimum age steps of the underlying SSP models, from $t$ = 0 to $t$ = 13.75 Gyr. The metallicity grid spans $\mathrm{\log_{10}(Z/\mathrm{Z_{\odot}})} = -2.2$ to $0.5$ with $\Delta Z = 0.3$ at low metallicity, with bin resolution increasing to $\Delta Z = 0.1$ near $Z = 0$. We use a \citet{chabrier2003} IMF throughout, and include the AGB dust emission model of \citet{villaume2015}. To cover the most recent star formation and avoid resolution effects of stellar particles, we set the star formation rates younger than 15 Myr equal to the instantaneous star formation rate from the galaxy's gas particles with metallicity equal to the current mass-weighted metallicity of the star-forming gas.
        
        We generate the spectrum of a simple stellar population at each point in the $t,Z$ grid. The sum of the SPS spectra, weighted by the stellar mass formed, produces the galaxy spectrum. The nebular emission lines are calculated independently for each stellar metallicity bin using the \texttt{FSPS} nebular emission line prescription based on a CLOUDY lookup table \citep{byler2017}, with the gas metallicity fixed to that of the metallicity bin. 
        
        Dust forms a crucial component to consider when building mock observables and when analyzing observational data, and different dust models can strongly affect results and conclusions. However, for lower-mass galaxies with relatively low gas masses and low metallicities, dust is expected to affect results less strongly. We therefore apply the well-known and often used two-component dust model from \citet{charlot2000}, which consists of a dust screen with a power law dust attenuation curve with index $\Gamma = -0.7$ and a normalization of $\tau(5500 \angstrom) = 0.33$. Stars younger than 30~Myr are additionally attenuated following an identical power-law attenuation curve with $\tau(5500 \angstrom) = 1.0$. We discuss the effects of using different dust models in building mock spectra in Starkenburg et al. (in prep.). 
        
        Hahn et al. (in prep.) use approximate Bayesian computation to infer an empirical dust model using the same set of mock galaxy spectra by fitting SDSS $r$-band magnitudes, $g-r$ colors, and NUV-FUV colors. We have tested how results in the present work change when using the best-fit dust empirical model. While the \ha\ EW measurements can be significantly affected by changing the dust model, \dnfour\ is relatively unchanged, as the index is not strongly sensitive to dust attenuation.
        
        Using the synthetic spectra, we generate SDSS \textit{g} and \textit{r} band magnitudes, as well as \dnfour\ and \ha\ EW for all galaxies (\autoref{fig:mag_d400}, top row). These quantities are noise-free and derived from spectra which encompass the total stellar light of each galaxy.
    
    \subsection{Mock surveys}\label{subsec:sightlines}
    
        In generating synthetic photometric and spectroscopic quantities for each simulated galaxy, we are much closer to observational analogues for each simulation. However, the distribution of absolute magnitude and \dnfour\ shown in the top row of \autoref{fig:mag_d400} for each simulation are not directly comparable to the corresponding distribution in SDSS (upper left), as the simulations are unaffected by observational noise and each sample contains every galaxy in the simulation volume. To accurately match observations, we need to create mock surveys of each simulation box\footnote{\texttt{Orchard}, our package for creating mock surveys, can be accessed at https://github.com/IQcollaboratory/orchard}. 
        
        Our method for creating mock surveys is as follows:
        
        \begin{enumerate}
            \item Place an ``observer'' at a random location 10 Mpc outside the simulation volume. This distance is selected to match the lower distance limit on galaxies in the NSA catalog from SDSS.
            
            \item Calculate apparent magnitudes, radial velocities, and projected distances for all galaxies as the observer would see them ``on sky''. For each galaxy, the radial velocity is the sum of the peculiar velocities along the observer's sightline and the recessional velocities given the distance between the galaxy and observer. 
            
            \item Convolve the synthetic spectra with the SDSS instrumental line spread profile (modeled as a Gaussian with $\sigma = 70$ km/s) and resample the spectra to match the SDSS wavelength resolution and coverage. 
            
            \item Add realistic spectral noise. The average signal to noise ratio (SNR) of SDSS spectra is dependent on galaxy color and apparent magnitude, and is also a function of wavelength. To reproduce the noise characteristics of SDSS, we bin galaxies from SDSS as a function of $g-r$ color and apparent \textit{r} magnitude. In each bin, we randomly draw 50 spectra and calculate the average SNR at each wavelength. In each mock observer's frame, noise is added to the synthetic spectra based on galaxy color, apparent magnitude and wavelength by drawing from a normal distribution $\sigma(\lambda,\ g-r,\ m_r)$ such that the SNR matches the SDSS model. For simulated galaxies which fall outside the populated regions of the SDSS color-magnitude diagram, we select the closest bin in color-magnitude space. 
            
            \item Calculate stellar masses from the $g-r$ colors, following the prescription of \citet{mao2020}. Notably, this can result in stellar mass estimates that appear lower than the resolution limits of the simulation.
            
            \item Remeasure \dnfour\ and \ha\ EW from the noise-added and instrumentally-broadened spectra. \autoref{fig:d4ha_sfr} shows the distribution of noise-added \dnfour\ and \ha\ EW for the simulations in three stellar mass bins, as compared to SDSS. We follow the definition of \citet{balogh1999} for \dnfour\ measurements and \citet{yan2006} for \ha\ EW measurements.
            
            \item Remove ``unobservable'' galaxies from the sample. Spectroscopy was only acquired for galaxies in SDSS above the limiting magnitude $m_r = 17.7$. To accurately match SDSS, simulated galaxies which have apparent magnitudes fainter than 17.7 along a particular sightline must be removed from the sample.
        \end{enumerate}
        
        In \autoref{fig:mag_d400}, we show effects of adding realistic noise, velocity resolution, and completeness cuts to isolated, $\mstel < 10^{10} \ \msun$ galaxies in each of the three simulations. The upper row shows absolute magnitude ($M_r$) and \dnfour\ calculated from the noise-free synthetic spectra, while the bottom row shows \dnfour\ measured from the noise-added spectra and the apparent magnitudes ($m_r$) as seen by the observer along a random sightline.
        
        The gray region in each panel in the bottom row of \autoref{fig:mag_d400} represents the regime in which galaxies are too faint to be selected for spectroscopic follow up with SDSS. Of the three simulations, SIMBA produces the fewest ``unobservable'' galaxies. This is likely driven by the fact the SIMBA does not resolve galaxies all the way down to $\mstel = 10^8 \ \msun$, leading to fewer faint galaxies which are then preferentially removed from the observed sample (see \autoref{subsec:resolution} for a more detailed discussion on the effects of resolution in each simulation). TNG has the faintest population of quiescent galaxies in absolute magnitude (at least in part because it extends to the lowest stellar masses) and correspondingly fewer quiescent galaxies from this simulation are observable with an SDSS-like survey.
        
        In \autoref{fig:d4ha_sfr}, we show the distribution of \dnfour\ and \ha\ EW in three stellar mass bins for the observations and for the simulations as measured from the noise-added spectra. We color-code the simulations by instantaneous SFR, highlighting the need for synthetic observations to select the analogous quiescent population from simulations. As shown in \citet{geha2012}, no quiescent galaxies are observed in the SDSS volume below $\mstel = 10^9 \ \msun$. 
        
        \begin{figure}
        \epsscale{1.2}
            \plotone{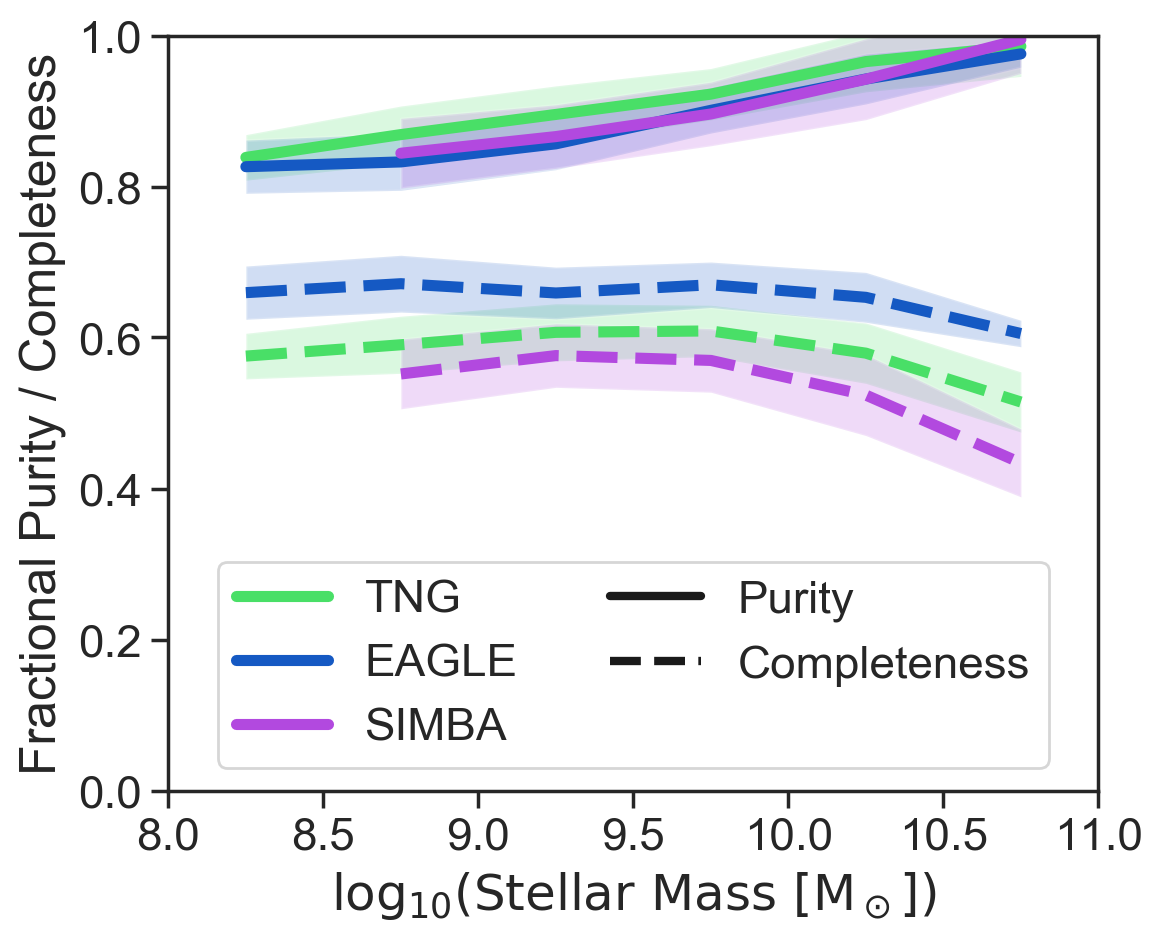}
            \caption{The purity (the fraction of galaxies selected as isolated by the \dhost\ criterion which are centrals, solid lines) and completeness (the fraction of the population of centrals which appear as isolated when using \dhost, dashed lines) of the \dhost\ isolation criteria compared to the central/satellite classification from the simulations themselves. Shaded regions represent the one-sigma variation across 10 randomly oriented sightlines. Using the \dhost\ criterion selects a sample of galaxies that is relatively pure ($> 85$ \% of isolated galaxies are centrals) but somewhat incomplete (only $55-70$ \% of centrals are selected as isolated). We use the true simulation-based stellar masses in this figure to provide maximum clarity on the purity and completeness of the \dhost\ criterion.}
            \label{fig:purity}
        \end{figure}

    \subsection{Isolation criteria}
    
        As in the observations, we select galaxies as isolated based on \dhost, the projected distance between each simulated galaxy and its most nearby massive neighbor. As in the observations, for each galaxy we identify potential hosts (galaxies with ($\mstel > 2.5 \times\ 10^{10}\ \msun$), within 7 Mpc in projected distance and 1000 km/s in radial velocity). For galaxies with $\mstel > 2.5 \times\ 10^{10}\ \msun$, we also require potential hosts to be more massive than the galaxy. Galaxies which have no potential hosts within 1.5 Mpc (\dhost $\ > 1.5$ Mpc) are considered to be isolated. 
        
        In \autoref{fig:purity} we compare our isolation criteria to the central/satellite classification from the simulations themselves. For all three simulation that classification is based on a halo finder algorithm that uses the underlying dark matter structure (not available in observations nor in our mock observations). The specific algorithm varies between EAGLE/TNG, and SIMBA. To define halos, both EAGLE and TNG use \texttt{SUBFIND} \citep{springel2001} to find overdense, gravitationally bound, (sub)structures within a larger connected structure that is found through a friends-of-friends \citep[FOF;][]{davis1985} group-size halo finder. In SIMBA,galaxies are identified as FOF groups of stars and star forming gas with spatial linking length of 0.0056 times the mean interparticle spacing, while halos are identified as FOF groups with linking length of 0.2 times the mean interparticle spacing. Galaxies and haloes are cross-matched in post-processing using the \texttt{yt}-based package \texttt{CAESAR}.
        
        For all three simulations the most massive subhalo (TNG and  EAGLE) or galaxy (SIMBA) in a larger group halo is then classified as the central galaxy, while all other subhalos are classified as satellites.
        
        For each simulation, we show the purity (the fraction of galaxies selected as isolated by the \dhost\ criterion which are centrals, solid lines) and completeness (the fraction of the population of centrals which appear as isolated when using \dhost, dashed lines) as a function of mass. For all simulations, purity declines with decreasing stellar mass, though all samples remain relatively pure even at low stellar mass. At $\mstel = 10^9 \ \msun$, a sample of isolated galaxies selected with \dhost\ will be $\sim 85$ \% centrals ($\sim 15$ \%  of the ``observed'' isolated sample of galaxies are actually satellites as defined by the halo finder). 
        
        Completeness is not strongly dependent on stellar mass below $\mstel = 10^{10} \ \msun$, but decreases above this threshold. Completeness varies more between simulations, possibly due in part to the use of different halo finders. At $\mstel = 10^9 \ \msun$, $55-70$ \% of the centrals selected by the halo finder will be observed as isolated, reflecting the more restrictive nature of the \dhost\ criterion. Recreating the observational selection criteria for isolation is a critical step in comparing between the population of isolated, quiescent galaxies selected from observations and those generated in simulations (see also \autoref{fig:halofinder} and the discussion at the end of \autoref{subsec:obs_compare}).

\begin{figure*}
\epsscale{1.1}
    \plotone{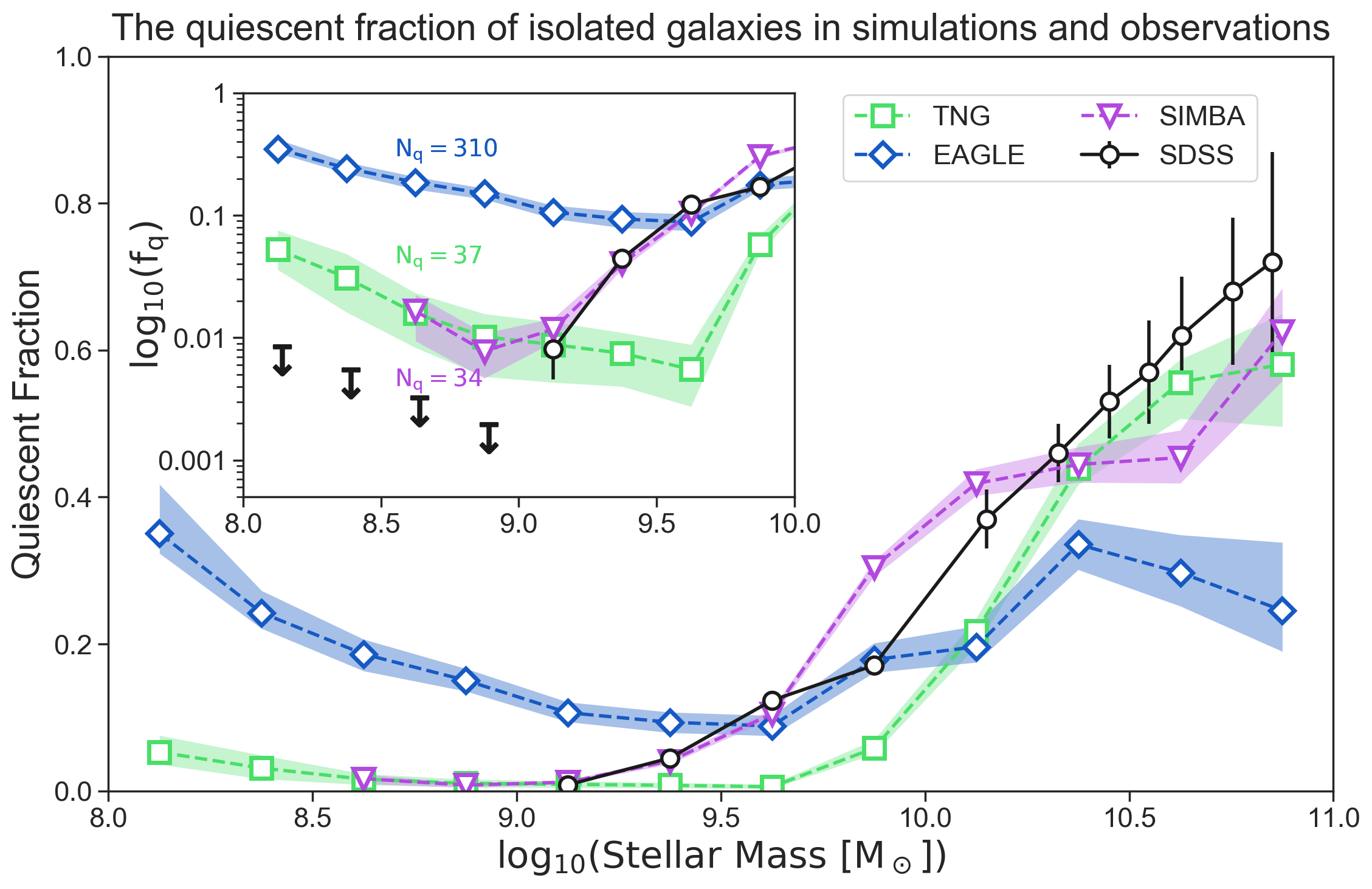}
    \caption{The median quiescent fractions of isolated galaxies as a function of stellar mass for SDSS (black circles), Illustris-TNG (green squares), EAGLE (blue diamonds), and SIMBA (purple triangles). The quiescent fractions for the simulations are the medians of 25 randomly placed sightlines around each simulation box. Shaded regions represent the combination of binomial uncertainty on the quiescent fraction and the variance across sightlines for each simulation. Errorbars on the SDSS quiescent fraction are adapted from \citet{geha2012}.
    \textbf{\textit{Inset:}} The same data shown in log-scale. Black arrows represent one-sigma upper limits for the SDSS data in bins where the number of isolated, quiescent galaxies is zero, and the number of isolated, quiescent galaxies ($\mathrm{N_q}$) in the lowest mass bin is indicated for each simulation.}
    \label{fig:qf_mass}
\end{figure*} 
        
    \subsection{Quenching criteria}
        
        As in the observations, simulated galaxies are considered to be quiescent if \ha\ EW $< 2$ \AA\ and \dnfour\ $> 0.6+0.1 \log_{10}(\mstel / \msun)$ (orange shaded region in \autoref{fig:d4ha_sfr}). Because \dnfour\ probes luminosity-weighted stellar age of a galaxy, it is difficult to make a corresponding selection in SFR space, as highlighted by the color-coding of galaxies in \autoref{fig:d4ha_sfr}. 
        
        The star forming and quiescent contours shown in \autoref{fig:mag_d400} can overlap because the \dnfour\ criterion is mass-dependent for each galaxy. Measuring \dnfour\ from the realistically-noisy synthetic spectra is crucial to accurately match the sample selection from observations. 
        
    \subsection{Volume correction}

        To calculate the quiescent fraction $f_q$, we weight each galaxy by the inverse of the total survey volume over which it could be observed given the SDSS spectroscopic magnitude limit ($1/\mathrm{V_{max}}$). 

        The quiescent fraction of isolated galaxies is then

        \begin{equation}
            f_{\mathrm{q}} = \frac{\sum\limits^{N_{\mathrm{q}}}_{i=1} 1 / V_{\mathrm{max},i}}{\sum\limits^{N_{\mathrm{q}}+N_{\mathrm{SF}}}_{i=1} 1 / V_{\mathrm{max},i}},
        \end{equation}
        
        where $\mathrm{N_{q}}$ and $\mathrm{N_{SF}}$ are the number of quiescent and star forming galaxies in isolation in each mass bin, respectively.
        
    \begin{figure}
    \epsscale{1.1}
        \plotone{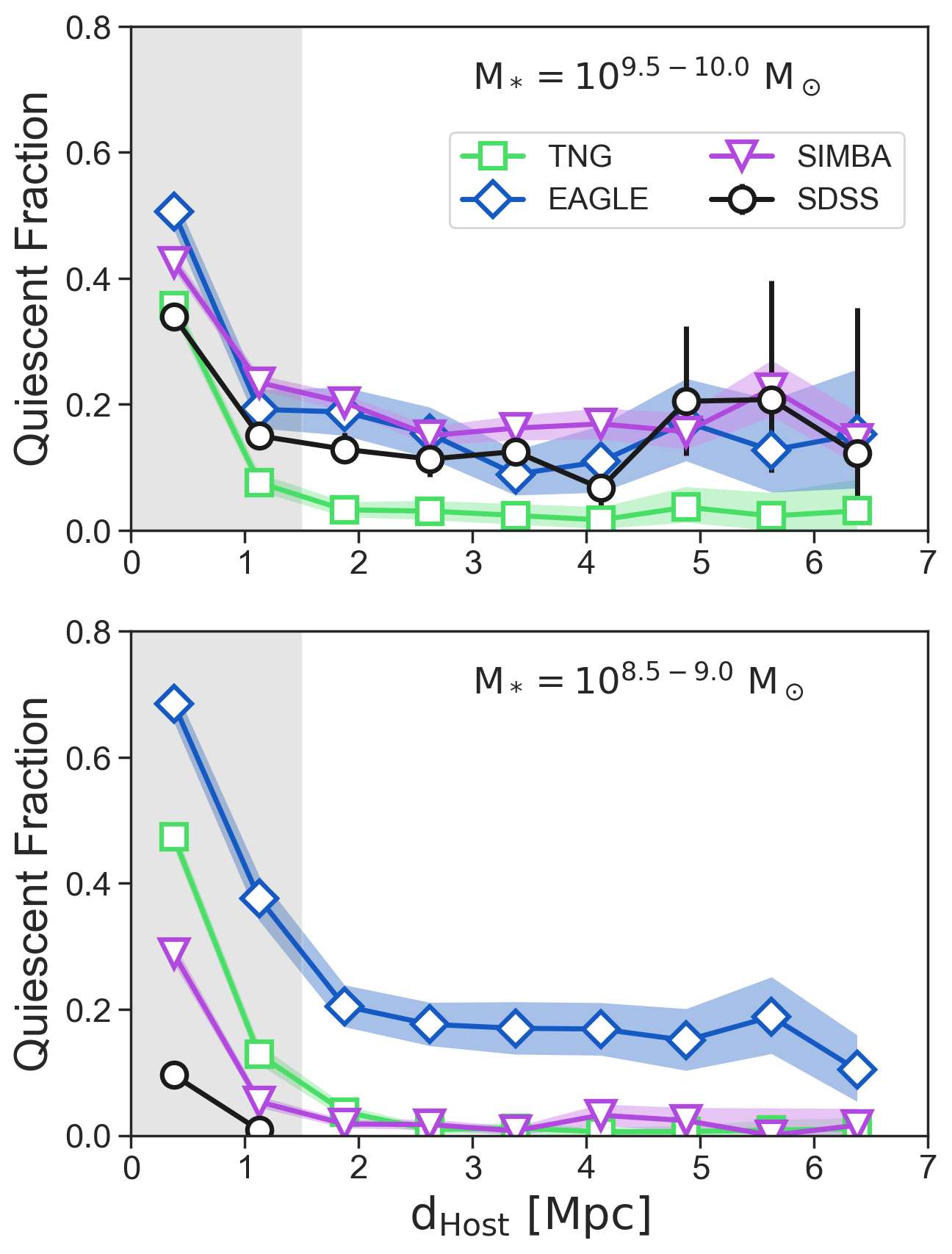}
        \caption{The quiescent fraction of galaxies as function of \dhost\ in two mass bins. The upper panel shows the observed distribution of galaxies with $\mstel = 10^{9.5-10.0} \ \msun$, and the lower panel shows $\mstel = 10^{8.5-9.0} \ \msun$. The gray region indicates \dhost $< 1.5$ Mpc, corresponding to non-isolated galaxies.}
        \label{fig:dhost}
    \end{figure} 
        
    \begin{figure}
    \epsscale{1.1}
        \plotone{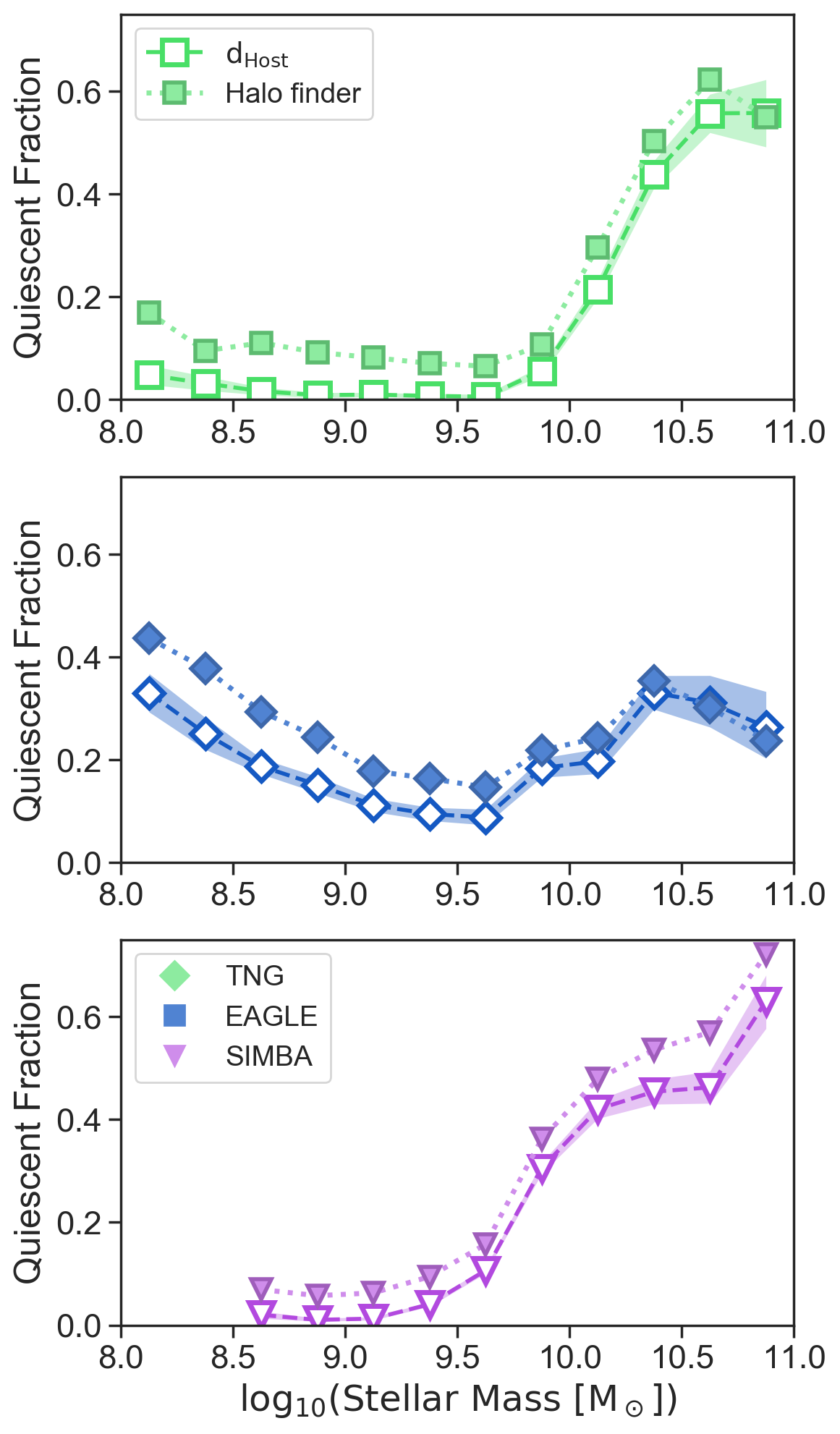}
        \caption{The effects of using a halo finder (filled-color) vs \dhost\ (open-face) to select isolated galaxies on the resultant quiescent fractions in TNG (top), EAGLE (middle), and SIMBA (lower panel). In all three simulations, using the halo finder leads to higher quiescent fractions when compared to \dhost, driven by the criterion's more restrictive nature.}
        \label{fig:halofinder}
    \end{figure} 

\section{The Quiescent Fraction of Isolated Galaxies}\label{sec:obs_v_sims}

In \autoref{subsec:obs_compare}, we compare the quiescent fraction of isolated galaxies from each simulation to observations from SDSS. In \autoref{subsec:resolution}, \autoref{subsec:mass} and \autoref{subsec:aperture} we discuss the effects of simulation resolution, observationally-biased stellar mass, and finite aperture on the quiescent fraction, respectively. 

\subsection{Comparing between simulations and observations}\label{subsec:obs_compare}
        
    In \autoref{fig:qf_mass}, we show the median isolated quiescent fractions for each simulation from 25 randomly placed sightlines (blue, green and purple points, dashed lines), as compared to SDSS (black circles, solid line). 
    
    At intermediate stellar masses ($\mstel = 10^{9.5-10.5} \ \msun$), the simulations all show the quiescent fraction decreasing with decreasing stellar mass, qualitatively reproducing the isolated quenching threshold seen in SDSS \citep{geha2012}. In observations, this is thought to be driven by the waning efficiency of internal feedback mechanisms. The threshold for TNG galaxies appears to be at a higher stellar mass than is seen in observations, while EAGLE's quiescent fraction appears to depend less strongly on stellar mass in this regime.
    
    Unlike the observations, all three simulations have non-zero quiescent fractions below $\msun = 10^{9} \ \mstel$ (\autoref{fig:qf_mass}, inset). For both SIMBA and TNG the quiescent fraction, though non-zero, remains small, with $f_q \sim 0.01$ at $\msun = 10^{9} \ \mstel$. Overproduction of quiescent galaxies in EAGLE is more pronounced. 

    We examine the distribution of quiescent galaxies as a function of environment in two mass bins just above and below the observed quenching threshold for isolated galaxies ($10^{8.5-9.0}$ and  $10^{9.5-10.0} \ \mstel$) in \autoref{fig:dhost}. At intermediate stellar masses (upper panel), EAGLE and SIMBA are a good qualitative match to observations, with low quiescent fractions for isolated galaxies and no environmental dependence in the isolated regime. TNG lies below the SDSS observations, with an average $f_q \sim 0.05$ beyond \dhost\ $= 1.5$ Mpc. At lower stellar masses (lower panel), EAGLE's over-production of quiescent galaxies in all environments is more pronounced, while SIMBA and TNG are in better agreement, with a near absence of quiescent galaxies beyond \dhost\ $= 1.5$ Mpc. 
    
    In general, the rapid increase of $f_q$ at \dhost\ $< 1.5$ Mpc is in qualitative agreement with observations. However, it is notable that all three simulations also overproduce low mass, \textit{non-isolated} quiescent galaxies, at a many sigma tension given the derived errors. Feedback models for satellite galaxies are often tuned to reproduce galaxies in the Local Group. The SAGA Survey \citep{mao2020} recently showed that the Local Group satellites may be over-quenched relative to those around more typical MW-mass galaxies, which may be driving this tension.
    
    In \autoref{fig:halofinder}, we show how the observed isolated quiescent fraction changes with different definitions of ``isolation''. For each simulation, we compare $f_q$ derived using \dhost\ and $f_q$ from each simulation's halo finder. In each case, the halo finder produces a larger quiescent fraction (with the exception of $\mstel > 10^{10.5} \ \msun$ in TNG and EAGLE). This effect varies as a function of stellar mass (as well as halo finder used), and highlights the differences between selecting a sample of central galaxies versus isolated galaxies.
    
    \subsection{Resolution effects}\label{subsec:resolution}
    
        All three cosmological simulations used in this work have multiple runs at varying resolutions. SIMBA and EAGLE both have smaller boxes with higher resolution and we use these boxes to test our results specifically with respect to resolution. While we do not use the higher resolution version of TNG (TNG50) we note that the effect of resolution on the colors and color bimodality of galaxies is described in \citet{nelson2018}. They argue that the main effect of resolution on galaxy colors is in using purely the stellar particles to derive a star formation history. While we avoid this effect by including instantaneous SFR in the SFHs used to generate our spectra (to reflect the most recent star formation), future work should address explicitly the resolution convergence properties of TNG for the quiescent fractions of low mass galaxies.
        
    \begin{figure}
    \epsscale{1.1}
        \plotone{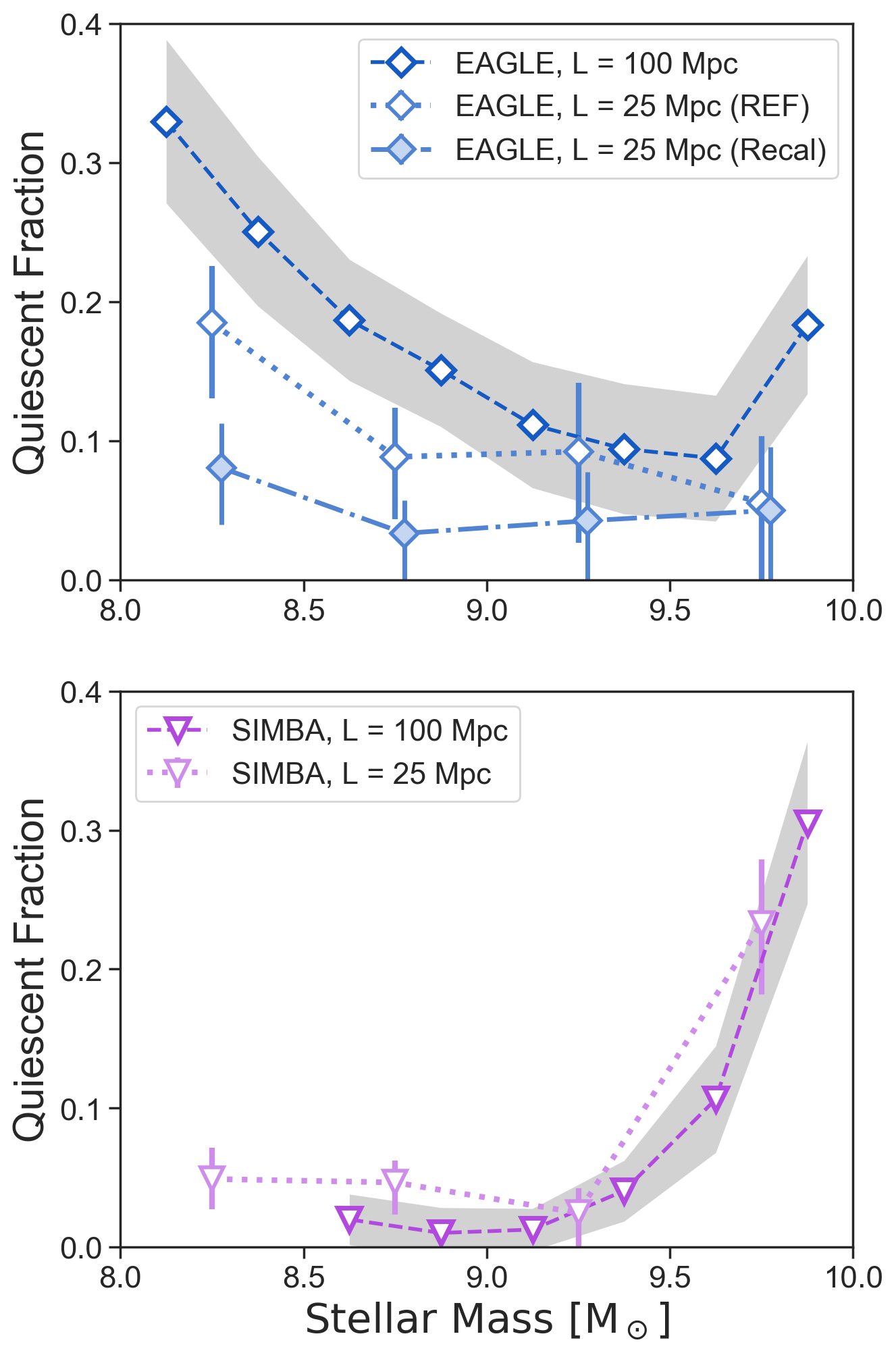}
        \caption{The effects of resolution on the isolated quiescent fraction in EAGLE (upper panel) and SIMBA (lower panel). In both panels, dashed lines show the fiducial quiescent fractions from the large volume, while dotted or dot-dashed lines show the quiescent fractions measured from the higher resolution (25 Mpc)$^3$ volumes. The gray shaded regions represent the maximum variation due to cosmic variance when considering (30 Mpc)$^3$ sub-volumes from each large volume simulation. In the upper panel, data from the two high resolution simulations are offset horizontally so that the errorbars may be distinguished.}
        \label{fig:resolution}
    \end{figure} 
        
        In \autoref{fig:resolution} we show the quiescent fraction as a function of stellar mass for the SIMBA and EAGLE higher resolution boxes (light colors, dot and dot-dashed lines) and compare those to the default box and resolution from \autoref{fig:qf_mass} (dark colors, dashed lines). Because the high-resolution boxes are much smaller in volume, we determine the effect from cosmic variance on the quiescent fraction. To do so, we recompute the quiescent fraction for subboxes of 30Mpc on a side from our default simulation boxes and show the full range of recovered quiescent fractions with the grey shaded regions in each panel. Due to the smaller total number of galaxies, we also use wider mass bins to calculate the quiescent fraction.
        
        For the EAGLE simulations we specifically compare to two boxes of 25Mpc on a side, run using the ``Reference'' (REF) version of the code (identical to the 100Mpc box), and using the ``Recal'' version of the code which is re-calibrated
        at this higher resolution of $m_{\rm DM} = 1.21 \times 10^6\ \msun$ and $m_{\rm gas} = 2.26 \times 10^5\ \msun$ to counterbalance resolution effects on the subgrid physics \citep[see][for a complete argument and description of the re-calibration, and a comparison of the different versions]{schaye2015, schaller2015}. 
        
        The quiescent fraction of the REF high-resolution box is slightly suppressed compared to that for the large box, falling below what can be explained by cosmic variance alone (gray shading) for the lowest masses. The re-calibrated high-resolution box (RECAL) are further suppressed relative to both the fiducial and REF boxes, and this effect is most extreme at the lowest stellar masses. \citet{schaye2015} argue that re-calibrating at different resolutions is most appropriate, as (un)physical effects of resolution may be hard to trace. Of the three EAGLE runs examined here, the RECAL box produces the fewest number of quiescent galaxies below $\mstel = 10^9 \ \msun$, and even agrees with a quiescent fraction of $f_q = 0$ within the uncertainties for three of the four stellar mass bins. The recalibrated high-resolution EAGLE box is therefore in closest agreement with the observations, though $f_q$ does not show the same strong dependence on stellar mass in the mass range $\mstel = 10^{9-10} \ \msun$ that is observed in SDSS. Additionally, both high resolution boxes still show an overabundance of low mass quiescent galaxies in isolation. 
        
        For SIMBA the quiescent fraction of low mass galaxies increases slightly at higher resolution, which is run with identical physics to the larger-scale box. We find that the high-resolution box agrees with the fiducial run, with a quiescent fraction that is slightly elevated relative to the fiducial run at $\mstel = 10^{8.5 - 9.0} \ \msun$. Improved resolution allows for the measurement of the quiescent fraction down to slightly lower stellar masses, which we find to be $f_q \sim 0.06$. This is still in tension with SDSS, where no quiescent galaxies are observed in isolation below $\mstel = 10^9 \ \msun$.
        
        We highlight that the effects of resolution, as exemplified by these case studies, are not uniform across simulations and can increase or decrease the quiescent galaxy fraction.

    \subsection{Stellar mass}\label{subsec:mass}
        In earlier work, \citet{munshi2013} found that the stellar masses measured for simulated galaxies with a combination of synthetic Petrosian magnitudes and \citet{bell2001} mass to light ($M/L$) ratios underestimate the true stellar mass by about 50\%. This underestimate varied only slightly as a function of mass. Similarly, \citet{leja2019} have shown that stellar masses inferred from SED modelling with non-parametric SFHs are roughly $\sim 0.2$ dex larger than those obtained under the assumption of an exponentially declining SFH. 
        
        We confirm the approximate magnitude of this effect using the $g-r$ color-based approximation for stellar mass from \citet{mao2020}, which produces stellar masses that underestimate the true stellar masses by $\sim 0.3$ dex. In addition to shifting the quenching threshold to lower stellar masses, accounting for systemic error in the measurement of stellar mass leads to a softening of the slope of the quiescent fraction.
    
    \subsection{Aperture effects}\label{subsec:aperture}
        Our fiducial mock galaxy spectra are based on galaxy properties and star formation histories calculated using all the star and gas elements considered to be part of a simulated galaxy's subhalo. In comparison, observations of galaxies in SDSS are aperture-limited. There are two SDSS apertures that are important for our results: the photometric aperture and the spectroscopic fiber aperture. The first roughly correspond to with what is considered the size of a galaxy and is relevant for the stellar masses used in this work, and the second for the \dnfour\ and \ha\ EW measurements. 
        
        With respect to ``galaxy size''-aperture effects: previous work using the EAGLE and TNG simulations have found that these aperture effects are significant for high-mass ($\mstel \gtrsim 10^{11}\ \msun$) galaxies but negligible for lower mass galaxies \citep{schaye2015,donnari2020_obs}, and we reach similar conclusions when comparing stellar masses in the full subhalo and within a 30 kpc aperture for TNG. Therefore, we conclude that aperture effects are likely insignificant for stellar masses, in particular when compared to other uncertainties as discussed in the previous section.
        
        The aperture of the SDSS spectroscopic fiber typically covers a few kpc in the central region of a galaxy. This means that aperture effects again, are crucial to take into account at higher masses, but can be small for low mass galaxies as there the SDSS fiber aperture may cover a significant fraction of the galaxy. We have checked the differences in \dnfour\ and \ha\ EW when only considering the central $r < 2$ kpc region of each galaxy. While \ha\ EW is more variable as a function of aperture (driven by changes in the amount of continuum contained within the aperture), galaxies do not significantly shift in or out of the quiescent region (which is based on both \ha\ EW and \dnfour). In particular, we find only small differences in the quiescent fractions for low mass galaxies, which shift upwards by $\sim 10\%$.
        
        For some observed galaxies the SDSS fiber may have been centered on an off-center bright (star-forming) region. With smaller (lower-mass) galaxies this possibility diminishes purely due to the fact that more of the total galaxy fits within the fiber aperture. Moreover, this is more likely to happen in actively star-forming galaxies, and as we purely use the spectral indices to classify galaxies as star-forming or quiescent, off-center fiber positioning is unlikely to affect our results.
        
\section{Quenching mechanisms in simulations}
    
\subsection{Comparison to previous work}\label{sec:disc}
    
        Our work is not the first to compare populations of quiescent galaxies from simulations to observations, though we are the first to compare three large volume simulations to observations in a fully consistent manner. Here, we review past studies evaluating the quiescent populations of each simulation.

    \subsubsection{EAGLE}\label{subsec:EAG_disc}
        The quiescent fraction of galaxies in the EAGLE simulations have been discussed in a multitude of papers, using a variety of galaxy subsamples, and of definitions and tracers of the quiescent fraction. \citet{schaye2015} find that the passive fraction for all galaxies defined based on specific star formation rates (sSFR; SFR/$M_\star$) is in reasonably good agreement with results from GAMA \citep{bauer2013} and SDSS \citep{moustakas2013}, although these observational results use slightly different criteria. The threshold where the galaxy population goes from predominantly blue and star-forming to predominantly red and quiescent is found to be at slightly higher stellar mass in EAGLE than for observed low-redshift datasets \citep{schaye2015, trayford2015}. \citet{furlong2015, trayford2015, trayford2017} show that the apparent larger quiescent fraction at the low mass end is largely due to resolution effects as low mass low-SFR galaxies have very few star-forming gas particles, and show that the recalibrated higher resolution box (Recal-25) agrees with observations. 
        
        We agree with their results that the recalibrated high-resolution box improves the quiescent fraction compared to the observations (see \autoref{fig:resolution} and \autoref{subsec:resolution}). Nevertheless, we still find that EAGLE produces a higher quiescent fraction for low mass galaxies than is observed in SDSS. It is noteworthy that EAGLE has a particularly high fraction of low mass galaxies that are classified as quiescent but do have current star formation at very low rates. This gives credence to the possibility that the SFR of these galaxies is poorly resolved and that these galaxies are misclassified as quiescent. However, our quiescent definition requires both low \ha\ and high \dnfour. While the \ha\ can be affected by these resolution effects, the \dnfour\ is based on the stellar continuum, and probes a longer timescale. Our quiescent fraction may therefore be less dependent on resolution than for a purely \ha-based definition.
        
        At higher masses the quiescent fraction we find for EAGLE seem particularly low compared to earlier work based on SFR or sSFR. However, \citet{trayford2017} show the similar results for \dnfour-defined quiescent fractions, and in addition show that dust can affect the discrepancy: when including dust the passive fraction in EAGLE based on a cut of \dnfour\ $> 1.8$ is 35\% reduced compared to results from SDSS for galaxies in the mass range $10^{10}\ M_{\odot} < M_* < 10^{11}\ M_{\odot}$, while without including dust this discrepancy increases to 70\% \citep{trayford2017}. These discrepancies are larger than what we find in this work and may be because our stellar mass-dependent \dnfour-based definition of quiescence, which is identical to the one used for the SDSS sample, reaches a lower value at similar masses.
    
    \subsubsection{Illustris-TNG}\label{subsec:TNG_disc}
        \citet{donnari2020_obs} and \citet{donnari2020} provide an in-depth exploration of the quenched galaxy fraction in TNG, exploring the effects of aperture, quenching definition, SFR timescale, environmental mis-identification, and incompleteness on the quenched fraction. While they focus on galaxies with $\mstel > 10^9 \ \msun$ and use SFR or SFS-based definitions of quiescence, we nonetheless find their fiducial quiescent fraction to be in excellent qualitative agreement with our results, specifically the existence of a quenching threshold just above $\mstel = 10^{10} \ \msun$, a small population of isolated (central) galaxies which are quiescent below $\mstel = 10^{9.5} \ \msun$, and a rapid increase in the number of quiescent galaxies from $\mstel = 10^{10-11} \ \msun$.
    
        \citet{donnari2019} describe the galaxy star forming sequence and the quenched fraction using different definitions and tracers for TNG100. They find that both a UVJ-selected quenched fraction and a distance from the star forming sequence-selected quenched fraction agrees reasonably well with observations, although they use a different UVJ selection for TNG than for the observations they compare to \citep[from][]{muzzin2013}. 
        
        In a by eye comparison, our TNG quiescent fractions (defined based on \dnfour\ and \ha) are overall lower than the UVJ-and-SFS-based quenched fraction from \citet{donnari2019}, in particular at the lower-mass end. A similar difference can be found between the UVJ-selected \citep{muzzin2013} and \dnfour\ and \ha-selected \citep{geha2012} observed quiescent fractions at $M_{\star} < 10^{10}\ M_{\odot}$. As it is unlikely that low mass star-forming galaxies move into the UVJ selection region due to dust, this suggests that the these low mass systems are predominantly red, but still exhibit some low (relatively recent) star formation which can be traced by their \dnfour\ and/or \ha EW.

    \subsubsection{SIMBA}\label{subsec:SIMBA_disc}
        \citet{dave2019} have compared the specific star formation rates of SIMBA galaxies to observations from the GALEX-SDSS-WISE Legacy Catalog \citep{salim2016,salim2018} and found good agreement, which is also consistent with our findings of very good agreement between the quiescent fraction from SIMBA and that observed in SDSS.
        
        \citet{dave2019} also suggest that the few low mass quiescent galaxies are in fact satellites misclassified by the FOF-based halo finder. Additionally \citet{christiansen2019, borrow2020} show that jet feedback from AGN in SIMBA can influence large regions around the AGN host galaxy and can entrain gas and influence gas in nearby galaxies. The SIMBA 100~Mpc box contains a massive cluster ($\mstel \sim 10^{15} \ \msun$) and the high-resolution 25~Mpc box also has an anomalously large central halo. However, we confirm that in both boxes, the majority of the quiescent galaxies we select are at least 2~Mpc away from the cluster or most massive halo center, and most are more than 5~Mpc away. Our results suggest that there may be additional effects driving the non-zero quiescent fraction at low mass (see \autoref{fig:dhost}, lower panel).
        
        Low mass, quiescent galaxies in SIMBA are also found to be over-sized compared to their star forming counterparts \citep{dave2019}. However, this should lead to a higher fraction of quiescent galaxies removed from the survey due to the SDSS surface brightness limits. Fixing this issue would therefore only increase the population of low mass, quiescent galaxies in SIMBA.
        
\subsection{Quenching mechanisms of low mass, isolated galaxies in simulations}\label{subsec:sim_quenching}    

    Given the observed over-abundance of quiescent galaxies at low mass in the simulations, we consider what feedback mechanisms might drive quenching in these systems.
    
    \textbf{Black holes:} some of the low mass galaxies in our sample will contain black holes (almost all galaxies above $\mstel = 10^9 \ \msun$ in TNG and EAGLE, and above $\mstel = 10^{9.5} \ \msun$ in SIMBA). Feedback from central supermassive black holes is often thought to be able to quench galaxies, and possible important in keeping galaxies quiescent (e.g., \citealp{somerville2015}, recent discussions include \citealp{zhu2020, terrazas2016, terrazas2020}). However, in the models the black hole feedback often becomes effective only at certain black hole masses \citep[][]{mcalpine2017, bower2017, weinberger2018, thomas2019, terrazas2020, habouzit2020}, which are not reached in low mass galaxies. Moreover, not all of the low mass galaxies in these simulations will host black holes (especially at $\mstel < 10^9 \ \msun$), so while black hole feedback could play some role, it is unlikely to the exclusive driver the quenching of the lowest mass galaxies in each simulation.

    \textbf{Star formation feedback:} Feedback from star formation is generally thought to reduce the efficiency of galaxy formation at the lower mass end \citep{somerville2015, schaller2015, schaye2015, pillepich2018a, dave2019}. Outflows from star formation feedback can temporarily expel large amounts of gas and decrease the gas density of galaxies. This is commonly seen in higher resolution simulations of lower mass galaxies (halo masses $\lesssim 10^{10}\ M_{\odot}$ \citealp{wright2019, wheeler2015, wheeler2019, rey2020}). These effects may result in temporary quiescence, but it is unclear if supernovae are capable of removing enough gas to fully quench galaxies at $\mstel \sim 10^9 \ \msun$. Temporarily quiescent low mass galaxies may however, compose part of our quiescent low mass galaxy samples. In large-scale simulations used in this work, the effectiveness of stellar feedback is likely also affected by resolution at the lowest mass end, and the effects of feedback and resolution may be hard to disentangle.
    
    \textbf{Splashbacks:} galaxies moving through larger halos can have their gas stripped away, reducing star formation. Splashback galaxies can be found up to $\sim 3 \mathrm{R_{vir}}$ \citep[see e.g.,][]{diemer2020} from their true host galaxy, making misidentification possible. However, our \dhost\ isolation criterion is more restrictive than the halo finders (see \autoref{fig:halofinder}), making it unlikely that more than a small fraction of the isolated quenched galaxies observed in each simulation are splashbacks.
    
    \textbf{Outflows from nearby massive galaxies:} \citealp{borrow2020} and \citealp{wright2019} show that the gas of low mass galaxies (both satellites and low mass centrals) can be stripped or entrained by jets and AGN outflows from more massive halos. While many of the observed isolated galaxies lie many virial radii from any massive halos, this gas removal effect could contribute to the elevated quenched fractions seen in the simulations.
            
    \begin{figure*}
        \epsscale{1.2}
        \plotone{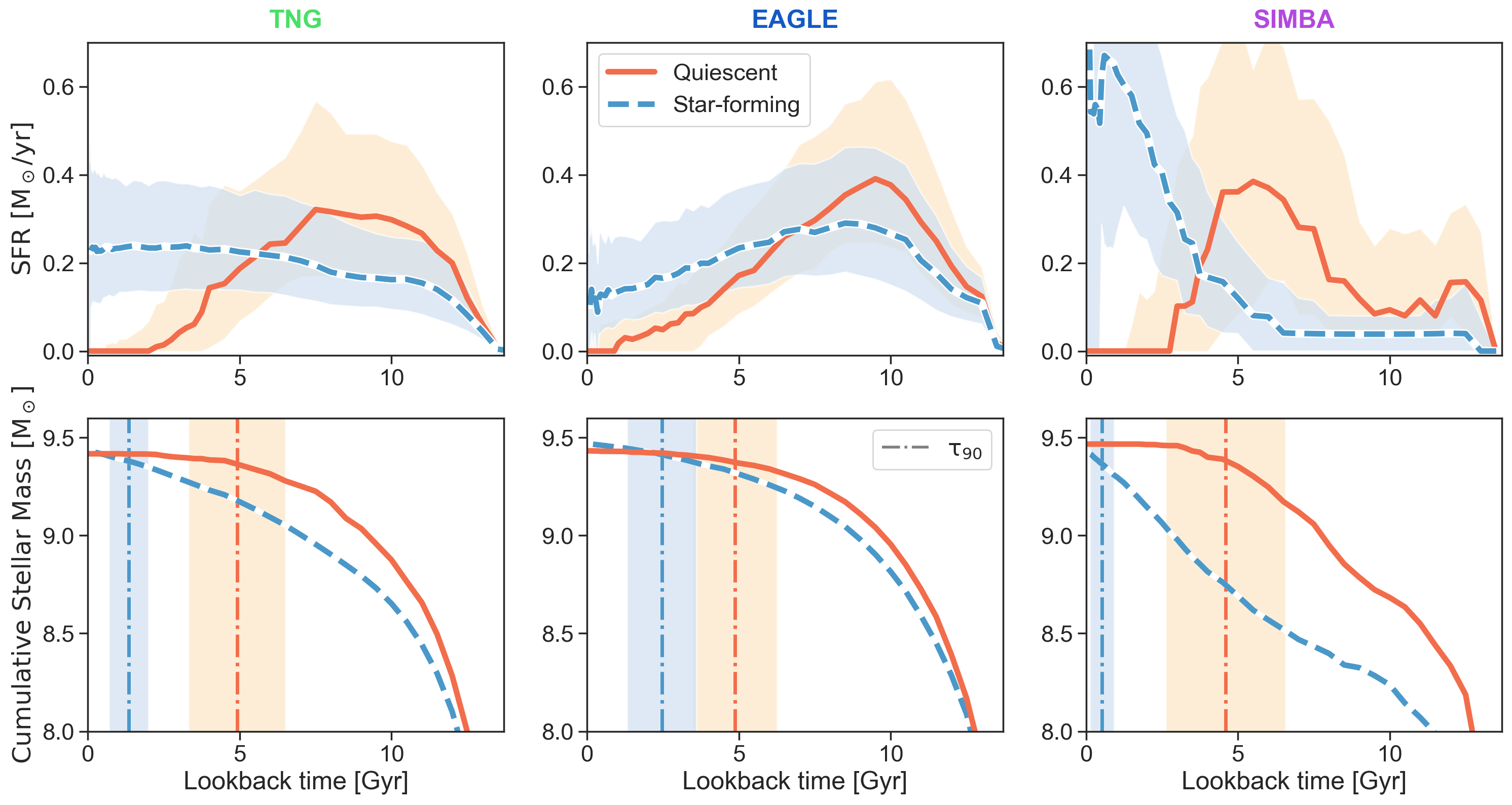}
        \caption{\textit{Top row:} The median star formation histories of low mass ($\mstel = 10^{9.0-9.5} \ \msun$), isolated galaxies in each simulation, split into star forming (blue, dashed) and quiescent (orange, solid), using the \dnfour-\ha\ EW criteria at $z = 0$. Shaded regions encompass the 20-80$^{th}$ percentile of each distribution. \\
        \textit{Bottom row:} The median cumulative stellar mass as a function of lookback time for the star forming and quiescent samples of isolated galaxies at low mass in each simulation. Vertical dot-dashed lines show the average time at which 90\% of a galaxy's stellar mass had formed ($\tau_{90}$). Though the star formation histories of quiescent galaxies vary across the three simulations, the average quenching timescales derived for the three populations are effectively identical ($\tau_{90} \sim 5$ Gyr).}
        \label{fig:sfhs}
    \end{figure*}
        
\section{Star Formation Histories of Quiescent Galaxies}\label{sec:SFH}

    By forward modelling simulated galaxies into observational space, we  gain the ability to look back at the ``true'' properties of observationally-selected galaxies. One of the most informative properties of a galaxy is its star formation history. For each simulated galaxy, we are able to compare the observed properties back to the biases inherent in the recovery of star formation histories from real data.
    
    In \autoref{fig:sfhs} (upper row), we show the median star formation histories of low mass ($\mstel = 10^{9.0-9.5} \ \msun$) isolated galaxies observed as star-forming (blue dashed lines) and quiescent (orange solid lines) at $z = 0$ in each simulation. The shape of the median star formation histories in each simulation varies significantly.  Low mass star forming galaxies in SIMBA show late-time rising star formation histories, while the same galaxies in TNG have approximately flat star formation histories, both in clear contrast with their quiescent counterparts.
    
    In EAGLE, the star formation histories are declining at late times for both the star forming and quiescent populations, and in fact a significant fraction of the low mass quiescent galaxies appear to be forming stars at very low rates at late times. This may be connected to the relatively low scatter in the EAGLE galaxy star-forming sequence \citep[see][]{IQ_paper1} and in the EAGLE star formation histories \citep{iyer2020}. In particular, the declining and low star formation rates for EAGLE's low mass galaxies may be connected to the relatively high quiescent fraction that we find using the \dnfour\ and \ha\ EW selection. \autoref{fig:d4ha_sfr} illustrates that of the three simulations in our sample, EAGLE has a notably large fraction of low mass, low-SFR galaxies that are borderline star-forming when considering their specific star formation rates, but have such low SFR that they are classified as quiescent when using the \dnfour\ and \ha\ EW measurements.
    
    In the bottom row of \autoref{fig:sfhs} we show the median cumulative stellar mass as a function of lookback time for the same star forming and quiescent samples. The vertical dot-dashed lines show the average time at which quiescent and star-forming galaxies formed 90\% of their stellar mass \citep{skillman2017,weisz2015}. Despite the apparent variation in the late time star formation histories of the quiescent populations observed in each simulation, $\tau_{90}$ is nearly identical: $\tau_{90,\mathrm{TNG}} = 4.9 \pm 1.6$ Gyr, $\tau_{90,\mathrm{EAGLE}} = 4.9 \pm 1.4$ Gyr, and $\tau_{90,\mathrm{SIMBA}} = 4.6 \pm 2.0$ Gyr. The measurement of $\tau_{90} \sim 5$ Gyr for low mass quiescent galaxies in isolation is a testable prediction for the observational sample and should provide insight into the timescale of feedback mechanisms which drive quenching in low mass galaxies.
    
    Of the galaxies observed as quiescent, a subset in each simulation have non-zero instantaneous star formation rates at $z = 0$ (approximately 50\% of low mass galaxies in TNG and EAGLE, and 20\% in SIMBA). The empirical definition of quiescence used throughout this work (based on \dnfour\ and \ha\ EW) selects galaxies with a range of evolutionary histories and $z=0$ properties. We find that the SFR=0 galaxies do not separate out cleanly in \dnfour \, -\, \ha\ EW space (highlighted in \autoref{fig:d4ha_sfr}). In order to select a totally quiescent sample of galaxies, additional probes of SFR would be required. Characterizing exactly how much ongoing star formation a galaxy can experience while still being selected by a given definition of quiescence will help inform the selection of appropriate quenching criteria (e.g., UVJ vs SFR vs \dnfour) going forward. 
    
    In future work, we will constrain the average quenching time for observed quiescent galaxies in isolation via SED fitting, and apply the same process to the synthetic spectra. Applying the same SED fitting methods to the synthetic data is critical, as the observed differences in the late time star formation history may only be recoverable above a certain threshold of data quality (e.g., spectrum SNR, photometric wavelength coverage).

\section{Conclusions}\label{sec:conc}

    In this paper, we have produced mock SDSS-like surveys from three large volume hydrodynamic simulations (Illustris-TNG, EAGLE, and SIMBA) from which we measured the quiescent fraction of isolated galaxies and compared back to extant constraints from the local universe \citep{geha2012}. We find that: 
    
    \begin{enumerate}
        \item The three simulations examined in this work, when transformed into observational space using an identical methodology, produce three different dependencies of the quiescent fraction on stellar mass. Above $\mstel = 10^{9.5} \ \msun$, all three simulations qualitatively reproduce the declining quiescent fraction with decreasing stellar mass observed in SDSS.
        \item All three simulations have non-zero quiescent fractions below $\mstel = 10^{9.0} \ \msun$, in contrast to observations of the SDSS volume. This suggests that current models of feedback in the low stellar mass regime are too efficient. 
        \item Our empirical definition of quiescence selects low mass galaxies with a range of star formation histories when viewed over longer (many Gyr) timescales. However, these populations all show similar quenching timescales ($\tau_{90} \sim 5$ Gyr), which can be compared back to observations. Understanding how sensitive a particular definition of quenching is to formation history can inform future population studies. 
        \item Measuring the quiescent fraction in a higher resolution box does not fully resolve the overabundance of quiescent galaxies below $\mstel = 10^{9.0} \ \msun$. In fact, improved resolution can lead to either a significant decrease or slight increase in the measured quiescent fraction, depending on the simulation.  While the low mass quiescent fraction is not converged in the large-scale simulations, the discrepancy with observations persists at higher resolutions.
    \end{enumerate}

    The low mass quenching threshold of isolated galaxies represents an observational boundary condition; a stellar mass regime where internal feedback mechanisms become ineffective. Observations of the isolated galaxy quiescent fraction provide a unique probe into the delicate balance of internal feedback mechanisms in low mass galaxies. Understanding how well or poorly modern simulations of galaxy evolution can reproduce this feature is a novel test of feedback prescriptions, but requires the creation of mock observations and surveys to enable appropriate comparisons between the observed universe and simulations.
    
    Our method for producing mock surveys from large volume hydrodynamic simulations can also be applied to zoom-in simulations and semi-analytic models, and adapted to match other surveys and observations. Future work will explore the star formation histories recovered from synthetic observations as compared to those derived from observations, as well as the observed gas properties of simulated galaxies.
    
    The constraining power of the observations we compare to are set by the size of the SDSS volume and the limiting magnitude and surface brightness of the survey. Future wide-field surveys such as the Vera Rubin Observatory's Large Synoptic Survey Telescope (LSST; \citealp{ivezic2019}) and the Dark Energy Survey Instrument (DESI; \citealp{desi}) have the potential to substantially improve our census of the local universe, providing new constraints on the population of low mass, quiescent galaxies in the field.
    
    Making direct comparisons between observations and simulations requires the careful translation from the simulation to observational frame (or vice versa). In doing so, we gain novel insights into the ways that feedback shapes the evolution of galaxies. 
    
\acknowledgements{
    We thank the Illustris collaboration and the Virgo Consortium for making their simulation data publicly available, and the SIMBA collaboration for sharing their data with us. The EAGLE and SIMBA simulations were performed using the DiRAC-2 facility at Durham, managed by the ICC, and the PRACE facility Curie based in France at TGCC, CEA, Bruy\`{e}res-le-Ch\^{a}tel. 
    
    This research was supported in part through the computational resources and staff contributions provided by the Quest high performance computing facility at Northwestern University, which is jointly supported by the Office of the Provost, the Office for Research, and Northwestern University Information Technology. The data used in this work were, in part, hosted on facilities supported by the Scientific Computing Core at the Flatiron Institute, a division of the Simons Foundation, and the analysis was largely done using those facilities.
    
    The IQ (Isolated \& Quiescent) Collaboratory thanks the Flatiron Institute for hosting the collaboratory and its meetings. The Flatiron Institute is supported by the Simons Foundation. CMD is supported by a Professor's Grant from the Howard Hughes Medical Institute (PI: Geha). NM acknowledges support from the Klauss Tschira Foundation through the HITS Yale Program in Astrophysics (HYPA).
    
    Funding for the Sloan Digital Sky 
Survey IV has been provided by the 
Alfred P. Sloan Foundation, the U.S. 
Department of Energy Office of 
Science, and the Participating 
Institutions. 

SDSS-IV acknowledges support and 
resources from the Center for High 
Performance Computing  at the 
University of Utah. The SDSS 
website is www.sdss.org.

SDSS-IV is managed by the 
Astrophysical Research Consortium 
for the Participating Institutions 
of the SDSS Collaboration including 
the Brazilian Participation Group, 
the Carnegie Institution for Science, 
Carnegie Mellon University, Center for 
Astrophysics | Harvard \& 
Smithsonian, the Chilean Participation 
Group, the French Participation Group, 
Instituto de Astrof\'isica de 
Canarias, The Johns Hopkins 
University, Kavli Institute for the 
Physics and Mathematics of the 
Universe (IPMU) / University of 
Tokyo, the Korean Participation Group, 
Lawrence Berkeley National Laboratory, 
Leibniz Institut f\"ur Astrophysik 
Potsdam (AIP),  Max-Planck-Institut 
f\"ur Astronomie (MPIA Heidelberg), 
Max-Planck-Institut f\"ur 
Astrophysik (MPA Garching), 
Max-Planck-Institut f\"ur 
Extraterrestrische Physik (MPE), 
National Astronomical Observatories of 
China, New Mexico State University, 
New York University, University of 
Notre Dame, Observat\'ario 
Nacional / MCTI, The Ohio State 
University, Pennsylvania State 
University, Shanghai 
Astronomical Observatory, United 
Kingdom Participation Group, 
Universidad Nacional Aut\'onoma 
de M\'exico, University of Arizona, 
University of Colorado Boulder, 
University of Oxford, University of 
Portsmouth, University of Utah, 
University of Virginia, University 
of Washington, University of 
Wisconsin, Vanderbilt University, 
and Yale University.
}
    
    \software{Astropy \citep{astropy:2013, astropy:2018}, FSPS \citep{conroy2009, conroy2010}, IPython \citep{PER-GRA:2007}, Matplotlib \citep{Hunter:2007}, NumPy \citep{2020NumPy-Array}, Python \citep{10.5555/1593511}, Python-FSPS \citep{2014zndo.....12157F}, SciPy \citep{virtanen2020}}

\bibliographystyle{yahapj}
\bibliography{references}

\end{document}